\documentclass[12pt,preprint]{aastex}

\slugcomment{submitted to ApJ.}

\shorttitle{ Helium settling and rotation-induced mixing }
\shortauthors{Sylvie Vauclair and Sylvie Th\'eado}

\begin{document}


\title{ On the coupling between helium settling and rotation-induced
mixing
in stellar radiative zones: I- Analytical approach }


\author{Sylvie Vauclair and Sylvie Th\'eado} 
\affil{Laboratoire d'Astrophysique, Observatoire Midi-Pyr\'en\'ees,
14 avenue Edouard Belin, 31400 Toulouse, France}



\begin{abstract}
In the presence of rotation-induced mixing, element diffusion still
occurs
in stellar radiative zones, 
although at a slower rate than in the case of a complete 
stability of the stellar gas.
As a consequence, helium settling leads to vertical $\displaystyle \mu
$-gradients which,
due to the meridional circulation, turn into horizontal fluctuations.
Up to now, the feed-back effect of this process on the rotation-induced
mixing
was currently
neglected in the computations of abundance variations in stellar
surfaces, or artificially reduced.
Here we analyse its consequences and derive an approximate analytical
solution 
in a quasi-stationary case. We also discuss the relative importance of
the various physical effects which influence the meridional circulation
velocity.
In a second paper (Th\'eado and Vauclair
2002a), we will 
present a complete 2D numerical simulation of this process while a third
paper (Th\'eado and Vauclair 2002b) will be devoted to special
applications to Pop I stars.

\end{abstract}




\section{Introduction}

The fundamental importance of element settling (also called microscopic
diffusion) inside stars was 
recognized by the pioneers of stellar structure computations
(see references in Vauclair and Vauclair 1982). A star is a
self-gravitational gaseous sphere, composed of all possible kinds of
different chemical elements, with different masses and atomic spectra.
Due to the pressure and thermal gradients and to the selective radiative
transfer, individual elements diffuse inside the star, one with respect
to the other, leading to a slow but effective restructuration. Mixing
processes like convection, turbulence, rotation-induced internal
motions, compete with microscopic diffusion and decrease its efficiency
according to the strength of the resulting homogeneization. When Von
Zeipel (1924) demonstrated that, 
due to centrifugal effects on the equipotentials of gravity,
rotating stars could not be in radiative equilibrium, he concluded that
the resulting thermal imbalance lead to meridional circulation which
could be the reason why the effects of microscopic diffusion were not
seen in most stars.

Much later, when evidences of ``chemically peculiar stars" where
obtained, astrophysicists first discussed them in terms of nuclear or
spallation reactions. These explanations could not however account for
the observations (see, for example, Vauclair and Reeves 1972).
Meanwhile, the idea that these abundance anomalies could be due to
microscopic diffusion emerged again and gave interesting results
(Michaud 1970, Michaud et al. 1976 (MCV$^2$) , Vauclair
et al. 1978 a and b (V$^2$M and V$^2$SM)). From then on, for more than
two decades, this became ``the diffusion hypothesis" which was supposed
to be added to the physics of peculiar stars, for the aim of explaining
their anomalies. This interpretation was misleading in the sense that it
presented element diffusion as a somewhat ``ad hoc" process added in
some cases, forgetting its fundamental character in the structure of
self-gravitational spheres. 

The importance of element settling in the stellar evolution process has
recently been proved in the Sun by helioseismic investigations. It was
known for a long time that helium and metals should have diffused by
about $20\%$ down from the solar convective zone since the birth of the
Sun up to know (Aller and Chapman 1960). Recent comparisons between the
sound velocity inside the Sun as computed in the models and as deduced
from the inversion of seismic modes, confirmed that this settling really
occurred.

In fact, element diffusion is always present in stars, unless
hydrodynamical processes increase its time scale in such a way that it
becomes much larger than the stellar lifetime. In this framework, the
basic difficulty is not to account for the occurrence of abundance
anomalies in some stars, but to understand why the consequences of
diffusion processes are not seen in all the stars and also why, in
chemically peculiar stars, the observed anomalies are not as strong as
predicted by the theory. This is related to the hydrodynamical processes
which occur in stellar radiative zones and may slow down the element
settling. Rotation-induced mixing is certainly one of the most important
of these processes.

The process of meridional circulation and its consequences
have been studied by many authors, including Eddington (1926), 
Sweet (1950), Mestel (1953, 1957, 1961), Tassoul and Tassoul
(1982, 1983, 1989), Zahn (1992, 1993), Maeder and Zahn (1998). The way
such processes can slow
down diffusion have also been studied many times in the literature,
beginning with Schatzman (1977), V$^2$SM and V$^2$M. 

Zahn (1992, 1993) suggested that the transport of angular momentum 
induced by the meridional flow
could lead to shear flow instabilities. Due to the density
stratification,
these instabilities should create anisotropic turbulence, 
more important in the horizontal than in the vertical direction. The
coupling between the meridional advection and the horizontal turbulence
would lead to a special kind of mixing for the chemical species,
parametrized as
an effective diffusion coefficient. Such a description has been used,
for example, by Pinsonneault, Deliyannis and Demarque (1992),
Charbonnel, Vauclair and Zahn (1992), Richard et al (1996)(RVCD). 

Later on, the transport of chemical species and angular momentum were
consistently coupled in the computations, with the assumption that no
other process interferred with rotation-induced mixing (Zahn, Talon and
Matias 1997, Talon and Charbonnel 1998).The results 
predicted a non negligible differential rotation inside the 
solar radiative regions, in contradiction with helioseismology.
The fact that the Sun rotates like a solid body below the convection
zone proves that angular momentum has to be tranported by another 
process, presumably magnetic fields (Mestel, Moss and Tayler 1988,
Charbonneau and Mac Gregor 1992, Gough and MacIntyre 1998).

>From the beginning of these computations of rotation-induced mixing, the
importance of the feed-back effect due nuclearly-induced $\displaystyle
\mu $-gradients was well recognized. Mestel and Moss (1986) showed how
these $\displaystyle \mu $-gradients could slowly stabilize the
circulation and expel it from the core towards the external layers. They
called this process ``creeping paralysis". Indeed, without such a
stabilization of the meridional circulation, the stars could not become
red giants in the correct time scale.

However, in all these computations the feed-back effect due to the
diffusion-induced $\displaystyle \mu $-gradients on the meridional
circulation was not included or artificially reduced to prevent
numerical instabilities. Vauclair (1999) showed how, in slowly rotating
stars, the resulting terms in the computations of the circulation
velocity could rapidly become of the same order as the other terms.
Th\'eado and Vauclair (2001) computed these terms for the case of Pop II
stars and claimed that the induced diffusion-circulation coupling could
be the reason for the very small dispersion of the lithium abundances
observed in halo stars.

While the $\displaystyle \mu $ terms due to element settling have
definitely to be taken into account in the computations of
rotation-induced mixing, the details of the coupling process which
occurs when these terms become of the same order of magnitude as the
classical terms remain difficult to handle.
In the present paper, we introduce an analytical approach of the process
in solar type stars and discuss an approximate solution in a
quasi-stationary case. In
a second paper (Th\'eado and Vauclair 2002a, herafter referred to as
paper II), we will present the results of a 2D numerical simulation
which will help visualize the situation and give a prescription for
introducing this process in 1D stellar evolution codes. The case of
earlier spectral types, when the so-called ``Gratton-$\rm{\ddot{O}}$pik term"
becomes
larger than unity below the convective zone, will be discussed in a
third paper, with applications to galactic clusters (Th\'eado and
Vauclair 2002b, paper III). 

Our conclusion will be that the feed-back effect due to
diffusion-induced $\displaystyle \mu $-gradients may strongly modify the
meridional circulation and meanwhile reduce the efficiency of diffusion
as suggested in Th\'eado and Vauclair 2001.

\section{Rotation-induced mixing}

In rotating stars, due to the centrifugal forces, radiative equilibrium
is
not satisfied. This has to be compensated by a motion of matter with a
vertical velocity derived form the equation:
\begin{equation}
\rho T\left( \frac{\partial s}{\partial t}+\mathbf{u.\nabla }%
s\right) =-\mathbf{\nabla .}F+\rho \varepsilon _{n}
\end{equation}
where $\varepsilon _{n}$ is the nuclear energy production, negligible in
the
outer layers, F the radiative flux, s the entropy density, $\rho$ the local density, $T$ the local temperature, $u$ the matter velocity. 

>From eq.(1) we deduce the vertical component of the circulation velocity
which, if developped on the second Legendre polynomial, may be written
(Zahn 1993, Maeder and Zahn 1998, herafter MZ98):
\begin{equation}
u_{r}=U_{r}P_{2}(\cos \theta )
\end{equation}

Following Zahn 1993 and MZ98, all physical parameters are developped on a level surface in the form :
\begin{equation}
x = \overline{x} + \widetilde{x} P_{2}(\cos \theta )
\end{equation}

Then $U_{r}$ is obtained as :
\begin{equation}
U_{r}=\frac{P}{\overline{\rho g} C_{P}\overline{T}\left( \nabla
_{ad}-\nabla +\nabla _{\mu }\right) }\left[ \frac{L}{M_{\ast }}\left(
E_{\Omega
}+E_{\mu }\right) +{\overline{T}}C_{P}\frac{\partial \zeta }{\partial
t} \right] 
\end{equation}
where $P$ is the local pressure, $g$ the gravity, $ C_{P}$ the specific heat at constant pressure, and :
\begin{equation}
\zeta =\frac{\widetilde{\rho }}{\rho }=\frac{1}{3}\frac{r^{2}}{
\overline{g}}\frac{d\Omega ^{2}}{dr} \ 
\end{equation}

This parameter was referred to as $\Theta$ in MZ98 : we prefer to call
it $\zeta$ to avoid any confusion with the latitude angle, specially
when we treat 2D numerical simulations (paper II) ; note that, in MZ98
equation (4.38), $\overline{T}$ was forgotten in the second term of the
bracket, as well as in the first term of equation (4.36).

We have neglected the deviations from perfect gas law, which are not
important for the purpose of the present paper. The thermodynamical
parameters are averaged over level surfaces : in the following the bars
will be omitted for simplicity. Following MZ98, $L$ is the luminosity at
radius $r$ and
 $M_{\ast } = M(1-\frac{\Omega^{2}}{2\pi g\rho_{m}})$ where $M$ is the
stellar mass at radius $r$ and $\rho_{m}$ the mean density inside the
sphere of radius r. 

In most stellar situations, the term $\frac{\Omega^{2}}{2\pi g\rho_{m}}$
is negligible. In the present paper, we discuss the case of solar
type stars where the ``Gratton-$\ddot{O}$pik term" 
$\frac{\Omega^{2}}{2\pi g\rho}$
 (larger than the previous term as $\rho $ replaces $\rho_{m}$ in the
denominator) is always smaller than one below the convective zones.
Thus, for simplicity, we will replace $M_{\ast }$ by $M$ in the
following equations. The case where the ``Gratton-$\ddot{O}$pik term" is not
negligible in radiative zones (e.g. in F and A stars), will be studied
in a forthcoming paper (Th\'eado and Vauclair 2002b, paper III).

In equation (3), the terms related to the $\displaystyle \mu $-gradients
are gathered in $E_{\mu }$ while the classical terms appear in
$E_{\Omega}$. When the energy production terms are negligible (as in the
regions below the outer convective zone, when element settling is
important), $E_{\Omega}$ and $E_{\mu}$ are given by:
\begin{equation}
E_{\Omega }  =  \frac{8}{3}\left( \frac{\Omega ^{2}r^{3}}{GM}\right)
\left[ 1-\frac{\Omega ^{2}}{2\pi g\rho}\right] -\frac{\rho _{m}}{\rho
}\left[ 
\frac{r}{3}\frac{d}{dr}\left( H_{T}\frac{d\zeta }{dr}-\chi _{T} \zeta
\right) -\frac{2H_{T}}{r}\zeta +\frac{2}{3}\zeta \right] 
\end{equation}
and :
\begin{equation}
E_{\mu }  =   {\rho _{m} \over \rho  } \left\{{ r \over 3 } \
{d \over dr }\left[\left(H_{T} {d \Lambda \over dr}\right)\
- (\chi_{\mu } + \chi_{T} + 1) \Lambda \right]\
- {2 H_{T} \Lambda \over r } \right\}
\end{equation}
Here 
$\displaystyle H_{T}$ is the 
temperature scale height;
$\displaystyle \Lambda$  represents the 
horizontal $\displaystyle \mu $ fluctuations
$\displaystyle {\tilde{ \mu}\over \overline \mu } $;
$\displaystyle \chi _{\mu }$ and
$\displaystyle \chi _{T}$ represent the
derivatives:
\begin{equation}
\chi_{\mu } =
\left(
{\partial \ln \chi \over \partial \ln \mu  }\right)_{P,T}
\quad  ; \quad 
\chi_{T} =
\left( {\partial \ln \chi \over \partial \ln T }\right)_{P, \mu }
\end{equation}

>From equation (3), we find that three terms compete in the computation
of the meridional circulation velocity : $E_{\Omega }$, $E_{\mu }$ and
the third one, related to the density fluctuations, which we will write
: 
\begin{equation}
E_{\zeta }=\frac{M}{L} T C_{p}\frac{\partial \zeta }{\partial t}
\end{equation}

A fourth term has to be added in case of large horizontal turbulence, parametrized with a horizontal diffusion coefficient $D_{h}$. It
is due to the fact that such a turbulence modifies the radiative energy
transport. In MZ98, this term is added in the expression of $E_{\Omega
}$ (their equations 4.36 and 4.37). We prefer to add it separately as :
\begin{equation}
E_{h}=\frac{M }{L } \frac{6}{r^2}  C_p T D_{h} \zeta
\end{equation}
which can also be written :
\begin{equation}
E_{h}=\frac{\rho _{m}}{\rho } \frac{2H_{T}}{r} \frac{D_{h}}{K} \zeta
\end{equation}
where $K$ is the thermal diffusivity (Zahn 1993).
In this way, equation (3) may be written :
\begin{equation}
U_{r}=\frac{P}{\rho gTc_{P}\left( \nabla _{ad}-\nabla +\nabla _{\mu
}\right) } \frac{L}{M} E_{tot}
\end{equation}
with : 
\begin{equation}
 E_{tot} = E_{\Omega} + E_{\mu} + E_{\zeta } + E_{h}
\end{equation}

Note that the fourth term $E_{h}$ has a different behavior from the
other ones, as it is formally proportional to $U_{r}$ when the
horizontal
diffusion coefficient $D_{h}$ is approximated as in MZ98, that is $D_{h}
= C_{h} r \cdot U_{r}$ where $C_{h}$ is an unknown coefficient.
Rigorously,
it should be introduced in a different way, on the left side of equation
(11). We prefer to leave it presently on the right side and will show
later (section 3.3) how the prescription given for its parametrisation
induces that it must be negligible compared to the other terms.

Then the horizontal component of the meridional velocity is deduced from
the equation of matter conservation:
\begin{equation}
div\left( u_{r}\rho r^{2}\right) =-u_{\theta }\rho r
\end{equation}
so that :
\begin{equation}
u_{\theta }=-\frac{1}{2\rho r}\frac{d}{dr}\left( \rho r^{2}U_{r}\right)
\sin \theta \cos \theta 
\end{equation}

In the classical Eddington Sweet circulation, only $E_{\Omega }$ is
introduced in the computations. In this case, the divergence of the
upward flux is negative : for mass conservation, matter flows
horizontaly from the upward flux towards the downwards flux.
This situation may be dramatically modified when the other terms are
introduced, and particularly in the situation when $E_{\mu }$ becomes of
the same order as $E_{\Omega }$ while the two other terms are smaller.

\section{The importance of $\displaystyle \mu $-gradients }

According to the period in stellar evolution, the four terms included in
equation (11) may have various relative importance. Here we wish to
compare their orders of magnitude in solar type stars. Before that, we
have to discuss the various ways of computing the horizontal
$\displaystyle \mu $ gradients $\Lambda$ which enter the $E_{\mu }$
term.

\subsection {computations of $\displaystyle \Lambda$}

Due to the meridional circulation, vertical $\displaystyle \mu-$gradients 
give rise to horizontal ones as the upward flow brings up
matter with a larger $\displaystyle \mu $ while the downward flow brings
down matter with a smaller $\displaystyle \mu $. The resulting
horizontal $\displaystyle \mu $-gradients depend on whether horizontal
turbulence is present or not. 

In case of a laminar circulation (pure advection), where $u_{r}$ 
and $u_{\theta }$ are of the same order, the order
of magnitude of $\displaystyle \Lambda$ is simply given by:
\begin{equation}
\Lambda \simeq r \cdot \nabla \ln \mu 
\end{equation}
(the case where $u_{r}$ and $u_{\theta }$ have different orders of
magnitude will be came upon and studied in section 4) 

According to Zahn (1992) and Chaboyer and Zahn (1992),
the shears induced by the circulation lead to 
horizontal turbulence, which reduces the horizontal
$\displaystyle \mu $-gradients.
In this case,
introducing a  horizontal diffusion coefficient
$\displaystyle D_{h}$, the local variations of
$\displaystyle \mu $ are solutions of :
\begin{equation}
{\partial\tilde \mu  \over \partial t } +
u(r) \cdot
{\partial\overline \mu  \over dr } =
- {6 \over r^{2}}\ D_{h} \ \tilde \mu 
\end{equation}
In the stationary case, the horizontal
$\displaystyle \mu $-gradient including turbulence
is given by :
\begin{equation}
\Lambda = - {U_{r} \cdot r^{2} \over 6D_{h} } \
{\partial  \ln \mu \over \partial r }
\end{equation}
With the assumption that :
$D_{h} \simeq U_{r} \cdot r $, 
this equation becomes:
\begin{equation}
\Lambda \simeq - {1 \over 6 C_{h}}\
{\partial \ln \mu  \over \partial \ln r }
\end{equation}
where $C_{h}$ should be of order unity (MZ98, equation 4.41 ; 
note that we have chosen here the MZ98 definition of $C_{h}$, which is
the inverse of its definition in Chaboyer and Zahn 1992)

At the beginning of the stellar evolution on the main sequence,
$\displaystyle \Lambda$ is equal to zero. In the case of negligible
differential rotation, it increases and eventually reaches the critical
value for which $\displaystyle \vert E_{\mu }\vert \simeq 
\displaystyle \vert E_{\Omega}\vert$. 
Let us call this value 
$\Lambda _{crit}^{0}$. It only depends on the physical conditions in the
star, without any arbitrary parameter. If we neglect the derivatives of
$\Lambda$ (in paper II, all the terms will be included in the complete
2D simulation), we find :
\begin{equation}
\Lambda _{crit}^{0} = E_{\Omega }^{0} \frac{\rho}{\rho_{m}} \frac{r}{2
H_{T}}
\end{equation}
where $E_{\Omega }^{0}$ represents the part of the expression of
$E_{\Omega }$ which remains when the differential rotation is
negligible. 

We see that $\Lambda _{crit}^{0}$ varies like $\Omega ^{2}$. In solar
type stars, it is of order $10^{-6}$ for rotation velocities of order $5
$ km.s$^{-1}$ (Th\'eado and Vauclair 2001).
In case of non negligible differential rotation, the value of $\Lambda
_{crit}$ may become larger than $\Lambda _{crit}^{0}$. 

When there is a non negligible radial differential rotation in the
stellar radiative zone, $\zeta >0$. A special situation occurs if $\zeta$ becomes equal to 
$\Lambda $. The equation of state leads to :
\begin{equation}
\frac{\widetilde{\rho }}{\rho
}=-\frac{\widetilde{T}}{T}+\frac{\widetilde{\mu }}{\mu }
\end{equation}
or : 
\begin{equation}
\frac{\widetilde{T}}{T}=\Lambda -\zeta 
\end{equation}
(if the equation of state is different from perfect gas law, factors of
order unity have to be taken into account in these equations ; we
neglect them in the present discussion for simplicity)

If $\zeta =0$ , then $T/\mu $ is constant along a level surface. 
If $\zeta =\Lambda $, the temperature fluctuations vanish and 
the circulation stops unless forced by another process like angular momentum transport.
In this special case we may write $E_{\Omega }$ as :
\begin{equation}
E_{\Omega }=\frac{8}{3}\left( \frac{\Omega ^{2}r^{3}}{GM}\right) \left[
1-\frac{\Omega ^{2}}{2\pi g\rho}\right] -\frac{\rho _{m}}{\rho }\left[
\frac{r}{3}\frac{d}{dr}\left( H_{T}\frac{d\Lambda }{dr}-\chi _{T}\
\Lambda \right) -\frac{2H_{T}}{r}\Lambda +\frac{2}{3}\Lambda \right] 
\end{equation}

Then $\Lambda _{crit}$ becomes :
\begin{equation}
\Lambda _{crit} \simeq \frac{3E_{\Omega }^{0}}{2}
\end{equation}
which is about 100 times larger than $\Lambda _{crit}^{0}$ below the convective zone
of solar type stars.

\subsection {comparison between $E_{\Omega }$, $E_{\mu }$, $E_{\zeta }$
and $E_{h}$}

Suppose that $\zeta $ increases rapidly from $0$ to $\Lambda _{crit}$ at
the beginning of the stellar lifetime on the main sequence. 
Writing $\Lambda _{crit} \simeq \alpha E_{\Omega }^{0}$ we obtain :
\begin{equation}
E_{\zeta } \simeq \frac{M}{L} T\ C_{p} \alpha
\frac{d E_{\Omega }^{0}}{d t}
\end{equation}
As $E_{\Omega }^{0}$ varies like $\Omega^{2}$ , equation (26) may be
written :
\begin{equation}
E_{\zeta } \simeq \alpha   \frac{M T\ C_{p} }{L} 
E_{\Omega }^{0} \frac{1}{\Omega^{2}} 
\frac{d \Omega^{2}}{d t}
\end{equation}
so that $E_{\zeta }$ may be larger than $E_{\Omega }^{0}$ only if :
\begin{equation}
\frac{1}{\Omega^{2}} \frac{d \Omega^{2}}{d t} > 
\alpha \frac{L}{M T\ C_{p} }
E_{\Omega }^{0}
\end{equation}

In consequence, the term $E_{\zeta }$ may be dominant in equation (11)
during rotational braking, if the braking time scale is smaller than :

\begin{equation}
\Delta t < \alpha \frac{M T\ C_{p}}{L}
\end{equation}

For solar type stars, this time scale is of order $10^{14}$s, i.e. 
a few million years.

It is clear that the $E_{\zeta }$ term may be important only at the beginning of the
stellar lifetime, and only if the time scale for the angular momentum
transport is smaller than a few million years. Then, it becomes
negligible compared to the two other terms, $E_{\Omega }$ and $E_{\mu }$
.This situation may be different in late stages of stellar evolution, as
the internal structure of the stars then evolve in smaller time scales,
leading to a more rapid restructuration, while diffusion processes are
slowed down because of deep convective zone. However $\displaystyle \mu
$-barriers induced by nuclear reactions then have to be taken into
account, which may change the landscape.

Let us now discuss the $E_{h}$ term in equation (13), as given by
equation
(10).
This term is proportional to $\zeta$ and vanishes in case of negligible
differential rotation. Suppose that, on the contrary, this term becomes
preponderant in equation (12). Replacing $ D_{h} $ by its expression $
C_{h} r \cdot U_{r} $, this equation becomes :
\begin{equation}
U_{r}=\frac{P}{\overline{\rho g}\left( \nabla _{ad}-\nabla +\nabla _{\mu
}\right)} \frac{6}{r^2} C_{h} r U_{r} \zeta
\end{equation}
$U_{r}$ being on both sides disappears, and the coefficients in front of
$C_{h}\zeta$ are completely determined by the stellar structure, so that
the product of these two terms should be fixed, which does not seem
credible. At the bottom of convective zones in solar type stars, the
order of magnitude of the product $C_{h}\zeta$ as derived from equation 30 
is typically one.
As $\zeta$ is of order
$10^{-6}$ to $10^{-4}$ maximum, it would need a coefficient $C_{h}$ of
order
$10^4$ to $10^6$, much larger than one. 
In consequence, $E_{h}$ is always negligible in the
computations.

In the following section, we discuss the situation which happens when
$E_{\zeta }$ and $E_{h} $ are negligible and $\left| E_{\mu }\right|$
becomes of the same order as $\left| E_{\Omega }\right|$.

\section{Analytical solution in a quasi-stationary case}

Due to helium settling below the stellar convective zones in solar-type
stars, a $\displaystyle \mu $-gradient can built so that $\left| E_{\mu
}\right|$ becomes very close to $\left| E_{\Omega }\right|$ in a time
scale which depends on the local conditions but may be much shorter than
the main-sequence lifetime (Vauclair 1999, Th\'eado and Vauclair 2001).
In the case when $E_{\zeta } << E_{\Omega }$, a local process emerges
where too ``big terms" ($E_{\mu }$ and $E_{\Omega }$) nearly cancel each
other, thereby creating numerical instabilities in ordinary ``diffusion
and mixing" computations. In Vauclair 1999 and Th\'eado and Vauclair
2001, we suggested that both the mixing and the settling were then
strongly modified. Here we try to understand in more details what
physically happens in the star when this situation occurs. We study it
first in an approximate analytical way and later, in a second paper, we
will present a complete 2D numerical simulation of this process
(Th\'eado and
Vauclair 2002a, paper II).

If $E_{\zeta }$ is negligible, and $E_{\mu }$ nearly cancels $E_{\Omega
}$, we may expect, at first sight, that rotation-induced mixing freezes
out, like in the nuclear core of the star (Mestel and Moss 1986). This
``creeping paralysis" would first occur in the region just below the
convective zone, and slowly extend down to deeper layers. However such a
scheme forgets that helium settling would go on in this case, changing
both the vertical and horizontal $\displaystyle \mu $-gradients with
time. Thus no equilibrium could be reached, and circulation would
proceed again. This is the situation we analyse below.

Suppose that the circulation is nearly frozen in a region just below the
convective zone. This means that on every level surface, the horizontal
$\displaystyle \mu $-gradient $\displaystyle \Lambda$ is just equal to
the critical value $\displaystyle \Lambda_{crit}$, for which $\left|
E_{\mu }\right|$ is equal to $\left| E_{\Omega }\right|$. However,
helium diffuses out of the convective zone where it is completely
homogeneous, that is $\displaystyle \Lambda = 0 $. Because of this
diffusion, the horizontal $\displaystyle \mu $-gradient is always forced
to remain below the critical value in a boundary layer just below the
convective zone (in this analysis, we do not take into account any
overshooting or tachocline, but it would reinforce this conclusion). 
Under such conditions, the radiative unbalance which is responsible for
the meridional circulation and subsequent mixing still exists in this
boundary layer while it is mostly cancelled out in the frozen region below (but it still exists deeper in the star). 

Let us go back to the equation of matter conservation (14). In normal
(classical) meridional circulation, the flux divergence is negative and
matter is expelled horizontally from the upward flow to the downward
flow. Here, on the contrary, as the vertical circulation velocity
becomes 
very small below the convective zone except in the boundary layers where
it increases rapidly upwards,
the flux divergence is positive and matter is ``aspirated"
instead of being expelled. We expect that this situation must result in
a local loop of matter which closes horizontaly from the downward flow
to the upward flow (Figure 1). This may, in first approximation, lead to
a quasi-stationary stage that we discuss below.

\begin{figure}[h]
\epsscale{0.4}
\plotone{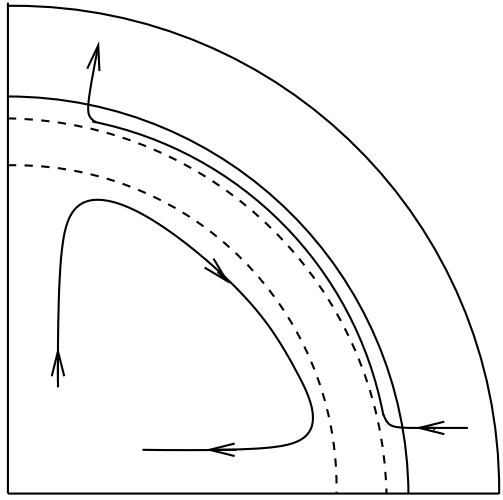}
\caption{Sketch of the circulation streamlines which may occur when the
two terms $E_{\mu }$. and $E_{\Omega }$ are of the same order of
magnitude. Normal meridional circulation goes on in the internal regions
where the diffusion-induced $\mu$-gradients are small (note that deeper
in the star the nuclearly-induced $\mu$-gradients should be taken into
account, which are not included here). In the ``frozen region", the
circulation is negligible except just below the convective zone where a
horizontal current is driven due to element settling. Exact streamlines
computed from a 2D numerical simulation will be given in paper II.}
\label{boucle}
\end{figure}

We assume that $\displaystyle\Lambda$ is equal to
$\displaystyle\Lambda_{crit}$ below the convective zone except in a few
layers where it decreases exponentially from 
$\displaystyle\Lambda_{crit}$ to 0. That is :
\begin{equation}
\Lambda \simeq \Lambda _{crit}\ \left[ 1-\exp \left( -\ \frac{r_{c}-\
r}{H}\right) \right] 
\end{equation}
where $H$ is a scale height which may be identified to the diffusion
scale height and $ r_{c}$ the radius at the bottom of the convective
zone.

Let us call $U_{\Omega }$ the value of $U_{r}$ obtained when $E_{\mu
}=0$ (classical meridional circulation). When $E_{\mu }\neq 0$ , we may
write :
\begin{equation}
U_{r}\simeq U_{\Omega }\left( 1+\frac{E_{\mu }}{E_{\Omega }}\right) 
\end{equation}
Assuming, in first approximation, that $E_{\mu }\propto \Lambda $,
equation (32) becomes:
\begin{equation}
U_{r}\simeq U_{\Omega }\left( 1-\frac{\Lambda }{\Lambda _{crit}}\right)
\simeq U_{\Omega }\exp \left( -\ \frac{r_{c}-\ r}{H}\right) 
\end{equation}

We may now calculate, from equation (14), the magnitude $U_{\theta }$ of
the horizontal
velocity which we define by : 
$u_{\theta }=U_{\theta}\sin \theta \cos \theta $

Neglecting the $\rho \left( r\right) $ variation in the boundary region, 
for a first approximation, we obtain :
\begin{equation}
U_{\theta }\simeq -U_{\Omega }\left[ 1+\frac{r}{2H}\right] \exp \left(
-\ \frac{r_{c}-\ r}{H}\right) 
\end{equation}
as $H<r$ , this equation becomes :
\begin{equation}
U_{\theta }\simeq -\frac{r}{2H}U_{\Omega }\exp \left( -\ \frac{r_{c}-\
r}{H}\right) 
\end{equation}
Thus the horizontal velocity is much larger that the vertical one for
matter conservation:

\begin{equation}
U_{\theta }>U_{r}
\end{equation}

So matter which settles down below the convective zone is swept up again
through this boundary loop. A quasi-stationary stage may take place for
which the horizontal $\mu -$gradient $\Lambda $ does not change with
time, except for the slow increase of the scale height $H.$

In this case, the continuity equation may be written in first
approximation :
\begin{equation}
U_{\theta } \frac{\partial \mu }{r \partial \theta} \simeq U_{r}  
\frac{\partial \mu }{\partial r}       
\end{equation}
so that  $\Lambda $ must be
written : 
\begin{equation}
\Lambda \simeq -\frac{U_{r}.r}{U_{\theta }}\frac{\partial \ln \mu
}{\partial r}\simeq 2H\ \frac{\partial \ln \mu }{\partial r}
\end{equation}
The vertical $\mu -$gradient is fixed from this relation :
\begin{equation}
\frac{\partial \ln \mu }{\partial r}\simeq \frac{\Lambda }{2H}\simeq
\frac{\Lambda _{crit}}{2H}\left[ 1-\exp \left( -\frac{r_{c}-r}{H}\right)
\right] 
\end{equation}

Although these computations are approximate, they show how the coupling
between helium settling and rotation-induced mixing can slow down both
the mixing and the settling when the $\mu $-terms are of the same order
as the $\Omega $-terms in the velocity of meridional circulation.

This does not completely stop the diffusion, but matter flows ajust
themselves so that, in the layers where this process occurs, the
vertical $\mu$-gradients remain constant. We see from figure 1 that
matter is flowing in opposite directions below the convective zone and
deeper in the star. This could create shear flow instabilities
in-between. However, as will be discussed in paper II, the frozen zone
is large enough and is the site of a stabilizing vertical $\mu
$-gradient, so that we expect this effect to be negligible.
As diffusion proceeds slowly,
the $H$ scale height increases with time : it is not a complete
stationary stage. But we expect the element settling to be significantly
reduced by this process, which may completely change the landscape of
abundance variations
in slowly rotating stars, as already suggested by Vauclair (1999) and
Th\'eado and Vauclair (2001).

\section{Discussion}

In the present paper, we have discussed the orders of magnitude of the
various terms which enter the computation of the meridional circulation
velocity in solar type stars. We have shown in an approximate way what
happens when the terms due to the diffusion-induced $\mu$-gradient
become of the same order as the classical terms in the circulation
velocity. In this case the pattern of the meridional circulation is
dramatically modified. 

When the $\mu$-terms are not taken into account (classical circulation),
a prescription may be used, as given by Zahn (1992), to compute the
transport of chemical elements. Due to angular momentum transport, a
shear flow instability may develop which leads to anisotropic
turbulence, smaller in the vertical than in the horizontal direction
because of the density gradient. The horizontal turbulence is
parametrised by a turbulent diffusion coefficient 
$D_{h} \simeq C_{h} r \cdot U_{r} $ while the vertical transport, which
is a combination of the meridional advection and the horizontal
turbulence, is treated with an effective diffusion coefficient :

\begin{equation}
D_{eff} \simeq \frac{ r U_{r} } { 30 C_{h} }  
\end{equation}

When the $\mu$-terms are taken into account, the prescription must be
modified. Horizontal turbulence has two effects : first it slows down
the construction of horizontal $\mu$-gradients, thereby delaying the
time when the $\mu$-terms  become of the same order of magnitude as the
$\Omega$-terms ; second it helps transporting energy horizontaly (the
so-called $E_{h}$ term in equation (12)). We have shown however that the
$E_{h}$ term can never be preponderant, because of its dependance on
$U_{r}$, which is consistent with the fact that $C_{h}$ should be of
order
one. In this case, as discussed in Vauclair and Th\'eado (2001) for halo
stars and will be discussed again for other cases (paper II and paper
III), the $\mu$-terms and the $\Omega$-terms become of the same order
during the main sequence lifetime of most solar type stars. Then the
process described here has to be taken into account.

We have shown that, when $\left| E_{\mu}\right|$ 
becomes very close to $\left| E_{\Omega }\right|$, matter flows may
ajust so that the vertical $\mu$-gradient $\nabla \ln \mu $ remains
constant in the ``frozen zone". This ``self-regulating process" will be
confirmed by the 2D numerical simulation (paper II). It can be used as a
prescription in stellar evolution codes for the computations of
abundance variations of chemical elements. It assumes however that the
time scale for restoring the equilibrium $\mu$-gradient is short
compared to the time scale of the element settling process which tends
to destabilize it. In other words, it assumes that $ U_{\Omega}$ is
larger than the microscopic diffusion velocity. This is the case below
the convective zone, but it may not be the case deeper in the star where
$ U_{\Omega}$ decreases quicker than the diffusion velocity. This is
why, in Vauclair and Th\'eado (2001), we assumed that the
``self-regulating " process occurred only in the region where 
$ U_{\Omega}$ was larger than $ V_{dif }$ and that microscopic diffusion
occurred unaltered below.

The question may arise of the orders of magnitude of the other terms in
equation (13) compared to the difference $\left| E_{\Omega }\right| -
\left| E_{\mu}\right|$ when this difference is small. In any case, as we
have seen, the term $E_{h }$ can never become preponderant. What about
the $E_{\zeta }$ term ?

$E_{\zeta }$ is given by equations (26) and (27). We know, from
rotational braking in solar type Pop I stars, that the external rotation
velocity decreases rapidly (in a few million years) from about
$100$km.s$^{-1}$ to a 
few km.s$^{-1}$ and then go on decreasing very slowly. The situation
inside the star depends on the process of angular momentum transport. We
know, from helioseismology, that a transport process more efficient than
the one induced by the circulation is needed for the angular momentum.
As discussed in section 3.3, $E_{\zeta }$ may be important at the
beginning of the stellar lifetime on the main-sequence, which is
consistent with the rotation braking time scale. For it to be also
important later, during the self-regulating process, it would need a
time scale (equation (29) modified) :

\begin{equation}
\Delta t < \frac{ E_{\Omega }}{ 
E_{\Omega } + E_{\mu}} \frac{M T\ C_{p}}{L}
\end{equation}

which can also be written :

\begin{equation}
\Delta t < ( 1-\frac{\Lambda }{\Lambda _{crit}} )^{-1} \frac{M
T\ C_{p}}{L}
\end{equation}

With $\Delta t$ of order $10^{17}$s ( a few billion years), equation
(41) can only be satisfied if 

\begin{equation}
( 1-\frac{\Lambda }{\Lambda _{crit}} ) < 10^{-3}
\end{equation}

which, in our exponential description, may occur only below a radius $r$
given by:

\begin{equation}
r_c - r \simeq 7 H
\end{equation}

In summary, mixing induced by the transport of angular momentum may be
important at the beginning of the main sequence stellar lifetime. Later
on, the situation below the convective zone is modulated by the
competition between the so-called $\Omega$-currents, due to the
classical meridional circulation and the so-called $\mu$-currents, due
to the diffusion-induced $\mu$-gradient.
In the present description, we have not taken into account other mixing processes which can decrease the horizontal chemical fluctuations, like tachocline layers. This should be added in future studies.
As proposed in Th\'eado and
Vauclair (2001), the reason why no large lithium depletion is observed
in most of Pop II main-sequence stars, in contradiction with Pop I
stars, could be due to a different rotation history : as the two stellar
populations were formed in the Galaxy in different circumstances and at
different ages, it could be that Pop II stars suffered less rotation
braking than Pop I stars so that the final lithium abundance and the
very small dispersion observed would be a consequence of the
``self-regulating process". 

This process, described here in an approximate physical way, will be
studied in paper II with a 2D numerical simulation.

\acknowledgments

We thank Jean-Paul Zahn for fruitful discussions, critics and comments.




\end{document}